\documentclass[]{aa}

\usepackage{txfonts}
\usepackage{graphicx}
\usepackage{amsmath}

\begin{document}

\title{Evolution of galaxy habitability}

\author{R. Gobat\inst{1}
\and S.E. Hong\inst{1}
}

\institute{
School of Physics, Korea Institute for Advanced Study, Hoegiro 85, Dongdaemun-gu, 
Seoul 02455, Republic of Korea
}

\date{}

\abstract{
We combine a semi-analytic model of galaxy evolution with constraints on circumstellar habitable zones 
and the distribution of terrestrial planets in order to probe the suitability of galaxies of different 
mass and type to host habitable planets, and how it evolves with time. We find that the fraction of 
stars with terrestrial planets in their habitable zone (known as habitability) depends only weakly on 
galaxy mass, with a maximum around $4\times10^{10}$~M$_{\odot}$. We estimate that 0.7\% of all stars in Milky 
Way-type galaxies to host a terrestrial planet within their habitable zone, consistent with the value 
derived from \emph{Kepler} observations. On the other hand, the habitability of passive galaxies is slightly 
but systematically higher, unless we assume an unrealistically high sensitivity of planets to supernovae. 
We find that the overall habitability of galaxies has not changed significantly in the last $\sim$8~Gyr, 
with most of the habitable planets in local disk galaxies having formed $\sim$1.5~Gyr before our own 
solar system. Finally, we expect that $\sim$1.4$\times10^9$~planets similar to present-day Earth have 
existed so far in our galaxy.
}


\titlerunning{The habitability of galaxies}
\authorrunning{Gobat \& Hong}

\maketitle

\section{Introduction}

The idea of a plurality of worlds in which the universe is filled with vast number of planets similar 
to our own has long been part of philosophical discourse and has acquired even more substance with the 
dawn of modern astronomy and the final shattering of the sphere of fixed stars. 
While Epicurus, Bruno or Herschel might have had little doubt that all these worlds were inhabited, 
our explorations of the sky have made us finally realize how unforgiving the cosmos could be for life, 
and we are beginning to wonder what other environments in the universe, if any, might actually support it. 

This question was partly answered in the last century with the concept of the circumstellar habitable zone 
\citep[HZ; e.g.,][]{Hua59,Kas93}, defined as the orbital belt around a star where the surface temperature 
of a planet would allow the existence of a biosphere. More recently, the idea of the HZ 
has by analogy been extended to that of the the galactic habitable zone \citep[GHZ;][]{Gon01}, i.e., the 
region within a galaxy where planets can form around stars and sustain a biosphere for a significant amount 
of time.
Initially mostly speculative, the GHZ has become increasingly quantifiable thanks to the data accumulated 
by exoplanet surveys during the last decade \citep[e.g.,][and references therein]{WG13}, as well as an 
increasingly better understanding of galaxy-scale physics. 
Modeling of the GHZ then tends to focus either on our own Milky Way \citep[MW; e.g.][]{Lin04,Pra08,Gow11} 
or specific nearby galaxies \citep[e.g.,][]{Spt14,For15}. The interest in our own surroundings is 
naturally high, and these studies benefit from the large amounts of data we have collected on them, 
allowing for fine-grained estimates. 
Generalizing then the GHZ to cosmic scales and epochs, we can discuss the capacity of the universe 
\citep[or even universe\emph{s}; e.g.,][]{Ad15} to sustain life up to the cosmic horizon, only 
limited by our understanding of galaxy evolution. 
This comes at the cost of simplification, however, either treating star-forming galaxies 
with (semi-)analytic recipes of galaxy evolution \citep{Day15,Zac16} or only considering the 
global star formation history (SFH) of the universe \citep{Lin01,Beh15}. 
This approach typically makes the simple assumption that the formation rate of habitable planets 
at any point in space and time is determined only by the local star formation rate (SFR) and metal 
content of the interstellar medium (ISM). 
Nevertheless, other factors can be expected to influence the number of habitable planets a galaxy 
can host, such as its stellar population, supernova rate (SNR), and structure. 
Moreover, early-type galaxies (ETGs), which are very common in the local universe, follow different 
scaling relations than star-forming disks and thus might have different habitability conditions. 

Here we take a semi-analytical approach that considers both star-forming and passive galaxies in a 
consistent way, as well as the effects of stellar evolution. This paper is structured as follows:
In Section \ref{method} we describe the model and its underlying assumptions, in Section 
\ref{results} we show the results under different initial parameters, while in Section \ref{conc} we 
summarize our conclusions.
\footnote{For reference, we assume a $\Lambda$CDM cosmology with $H_0=70$~km~s$^{-1}$~Mpc$^{-1}$, 
$\Omega_{\text{M}}=0.27$, and $\Omega_{\Lambda}=0.73$. However, the choice of cosmological 
parameters has little impact on our analysis or conclusions.}

\section{\label{method}Method}

In this section, we describe the different assumptions that underlie our estimates of galaxy habitability 
and its evolution with redshift. In particular, we rely on a treatment of galaxy evolution that includes 
the quenching of star formation. This is motivated by the predominance of passive galaxies in the 
local Universe \citep[e.g.,][]{F98}; while their structure and composition can not be directly 
equated with those of low star formation active systems, an analysis that considers both galaxy types 
requires that their evolution be approached in a consistent manner. 
Our objective here is not to recreate an accurate model of galaxy evolution and planetary formation, 
but to generate simple estimates that nevertheless self-consistently account for complexities 
(such as different stellar and galaxy types, that follow different evolutionary paths).

\subsection{\label{sfh}Galaxy evolution}

\begin{figure}
\centering
\includegraphics[width=0.45\textwidth]{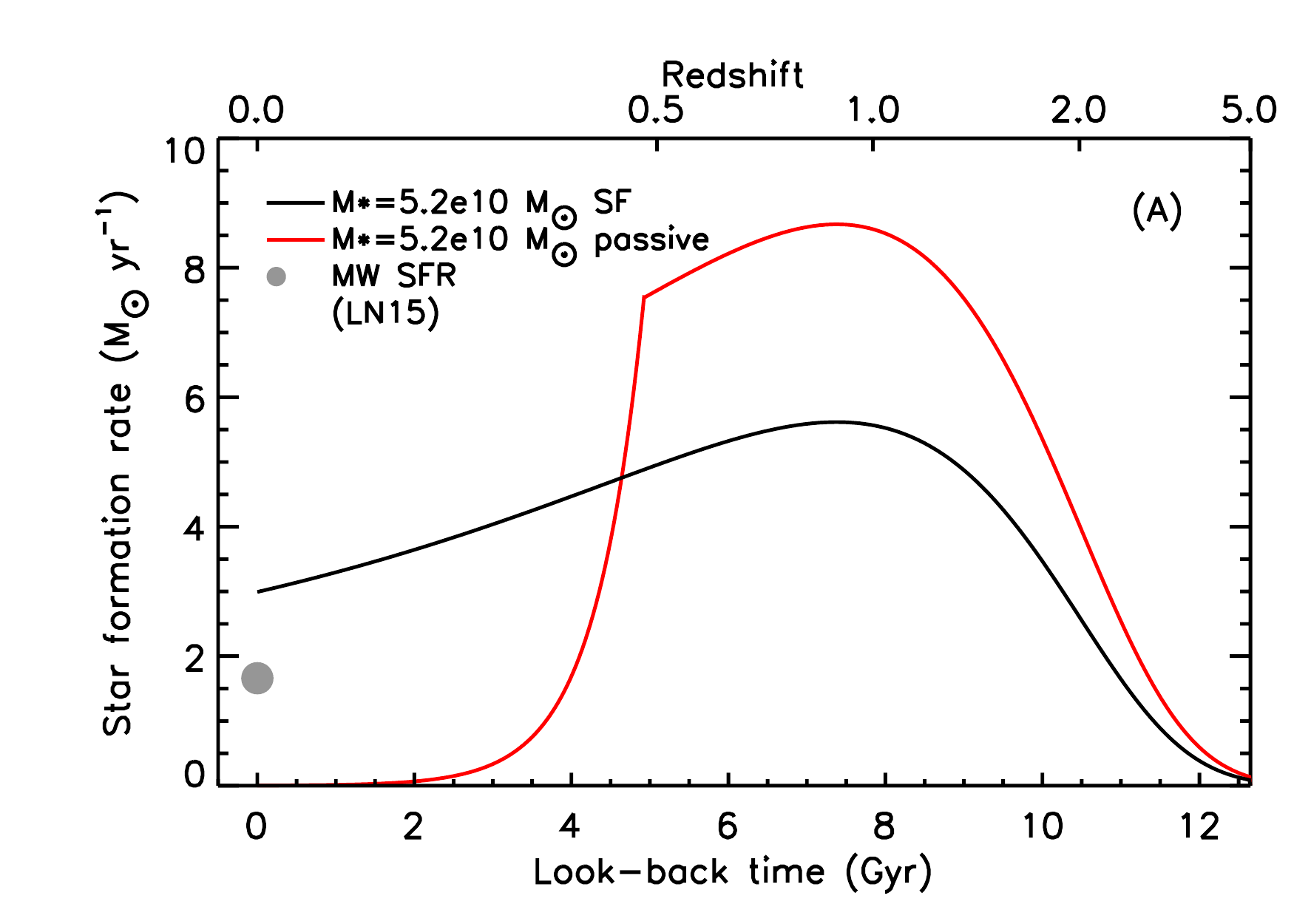}
\includegraphics[width=0.45\textwidth]{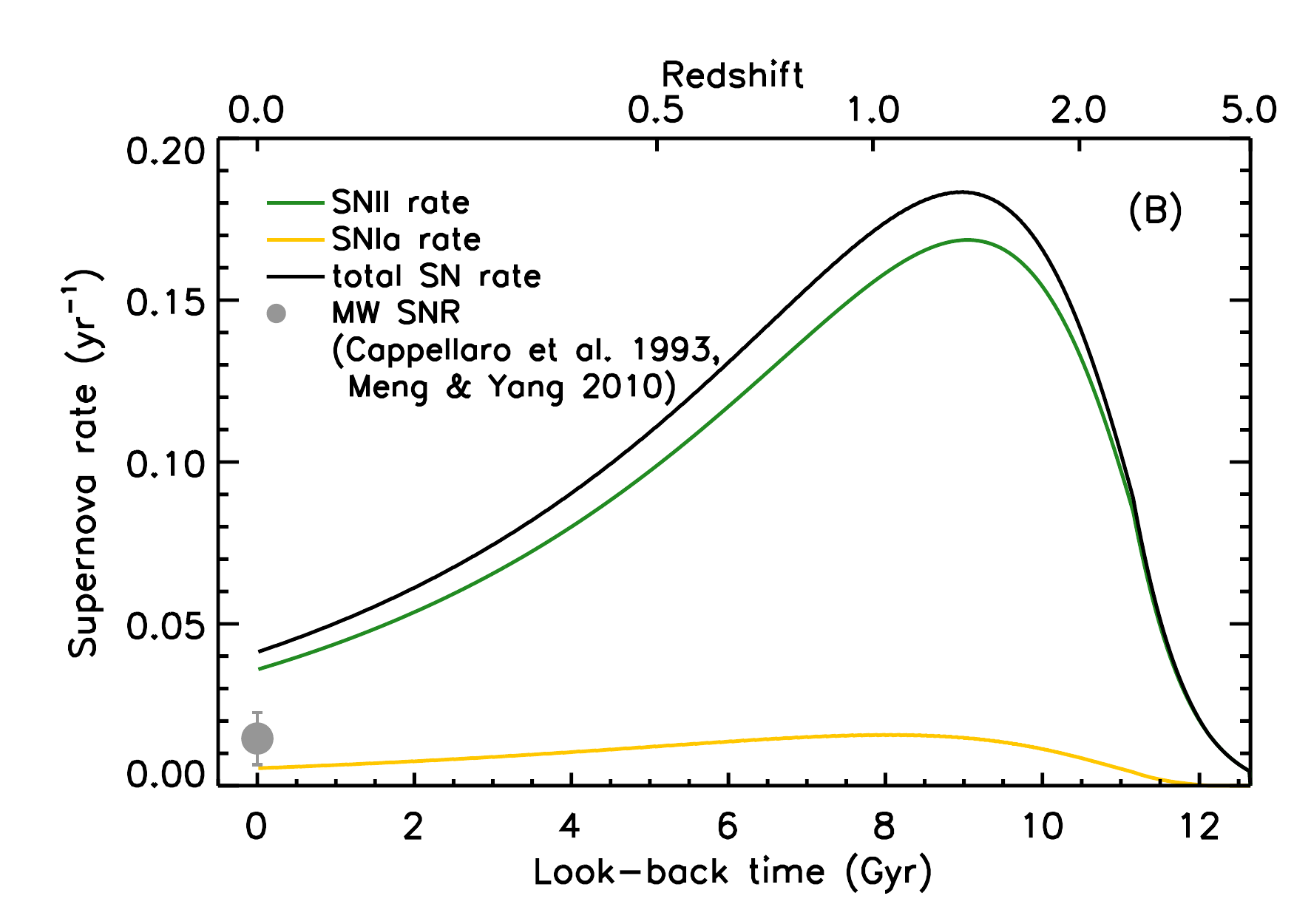}
\caption{\emph{(A):} SFHs of a passive galaxy quenching at $z=0.4$ (red) and a SF galaxy analogous to the 
MW (black), both having M$_{\star}=5.2\times10^{10}$~M$_{\odot}$. The gray dot shows the estimated SFR of 
the MW \citep{LN15}. \emph{(B):} calculated SNII (green), SNIa (yellow), and total SN rate (black) for the 
M$_{\star}=5.2\times10^{10}$~M$_{\odot}$~SF galaxy. The estimated MW value is shown as a gray dot \citep{Cpp93}.
}
\label{fig:sfh}
\end{figure}

We base our model of galaxy evolution on three main ingredients: a mass-dependent ``universal'' SFH, a 
single type of stellar initial mass function (IMF), and observed galaxy mass functions. 
We first assume that all galaxies start on the SFR - stellar mass relation 
\citep[SFR-M$_{\star}$, or ``main sequence'' of star formation; e.g.,][]{Bri04,Dad07,Rod10}, and describe the 
SFH of main sequence (MS) galaxies using the 2-SFM formalism \citep{Bet12,Sar14}. The slope and evolution of 
the SFR-M$_{\star}$~relation imply a SFH peaking at redshift $z=1-2$, with the SFR (here as a function of 
redshift) given by

\begin{equation}\label{eq:mssfh}
\Psi(z) = \text{sSFR}_{\text{MS},0} M_{\star}(z) 
\left(\frac{M_{\star}(z)}{10^{11}~\text{M}_{\odot}}\right)^{\alpha_{\text{MS}}-1} 
(1+\min(z,z_{\text{MS}}))^{\gamma_{\text{MS}}} ,
\end{equation}

\noindent
where $\gamma_{\text{MS}}=3$, $z_{\text{MS}}=2.5$, sSFR$_{\text{MS},0}=10^{-10.2}$~yr$^{-1}$~is the specific SFR 
(sSFR) of a $10^{11}$~M$_{\odot}$~galaxy at $z=0$, and $\alpha_{\text{MS}}=1$~is the slope of the 
SFR-M$_{\star}$~relation \citep{Ab14,Sch15}. The stellar mass M$_{\star}$~of the galaxy is then given by the 
time-integrated SFR minus losses due to stellar death. 
It is thus dependent on the choice of IMF (see Sect.~\ref{imf}). This SFH also requires a 
``seed'' mass as initial condition. Here we assume a single burst of star formation at $z_{\text{in}}=10$, with 
M$_{\star,\text{in}}=10^5-5\times10^9$~M$_{\odot}$. While the upper limit would yield an unrealistically high final 
galaxy mass at $z=0$ of $\sim$10$^{13}$~M$_{\odot}$, such a large seed mass is necessary in this context to generate 
the highly star-forming progenitors of massive $z\sim2$~passive galaxies (see Sect.~\ref{etg}). Examples of both a main 
sequence and quenched SFH are shown in Fig.~\ref{fig:sfh}. 
In addition, we assume that an evolving fraction $r_{\text{SB}}=0.012\times (1+\min(z,1))$~of star-forming galaxies 
are in starburst (SB) mode at any given time, with an average SFR enhancement of a factor 
$10^{0.6}$~\citep{Sar12,Bet12}. However, we here take the approximation that SBs do not contribute significantly 
to the M$_{\star}$~of the galaxy and only use the SB boost as a correction term to the SFR when estimating SNRs 
(Sect.~\ref{snr}). In this paper, all galaxies referred to as star-forming (SF) are either on the MS or are part of the 
SB fraction.

As long as a galaxy stays on the MS, we assume that its gas-phase metallicity $Z_{\text{g}}$~is given by the fundamental 
mass-metallicity relation \citep[FMR;][]{Man10}. A stellar population forming at time $t$ will then have a metallicity 
equal to $Z_{\text{g}}(t)$, with the average stellar metallicity of the galaxy $Z_{\star}$~being given by 

\begin{equation}\label{eq:Zstar}
\begin{aligned}
Z_{\star}(t) &= \frac{\int_0^t Z_{g}(t')\Psi(t')f_{\star}(t-t')\mathrm{d}t'}{\int_0^t \Psi(t')f_{\star}(t-t')\mathrm{d}t'}\\
f_{\star}(t) &= \int_{m_{\text{min}}}^{m_{\text{max}}} m\phi(m,t)\mathcal{H}(t_{\text{MS}}(m)-t) \mathrm{d}m\\
\end{aligned} ,
\end{equation}

\noindent
where $\mathcal{H}$~is the Heaviside function, $\phi(m,t)$ the IMF, and $t_{\text{MS}}(m)$ the stellar main sequence 
(sMS) life-time of a star of mass $m$ (Sect.~\ref{imf}); $m_{\text{min}}$ and $m_{\text{max}}$ are, respectively, the 
lower and upper limit on star masses. Here we choose $m_{\text{min}}=0.1$~M$_{\odot}$~(see Sec.~\ref{hab}) and 
$m_{\text{max}}=100$~M$_{\odot}$. The parameter $f_{\star}$~is then the remaining stellar mass fraction at time $t$, after 
accounting for losses due to stellar death. 

\subsubsection{\label{etg}Passive galaxies}

\begin{figure*}
\centering
\includegraphics[width=0.45\textwidth]{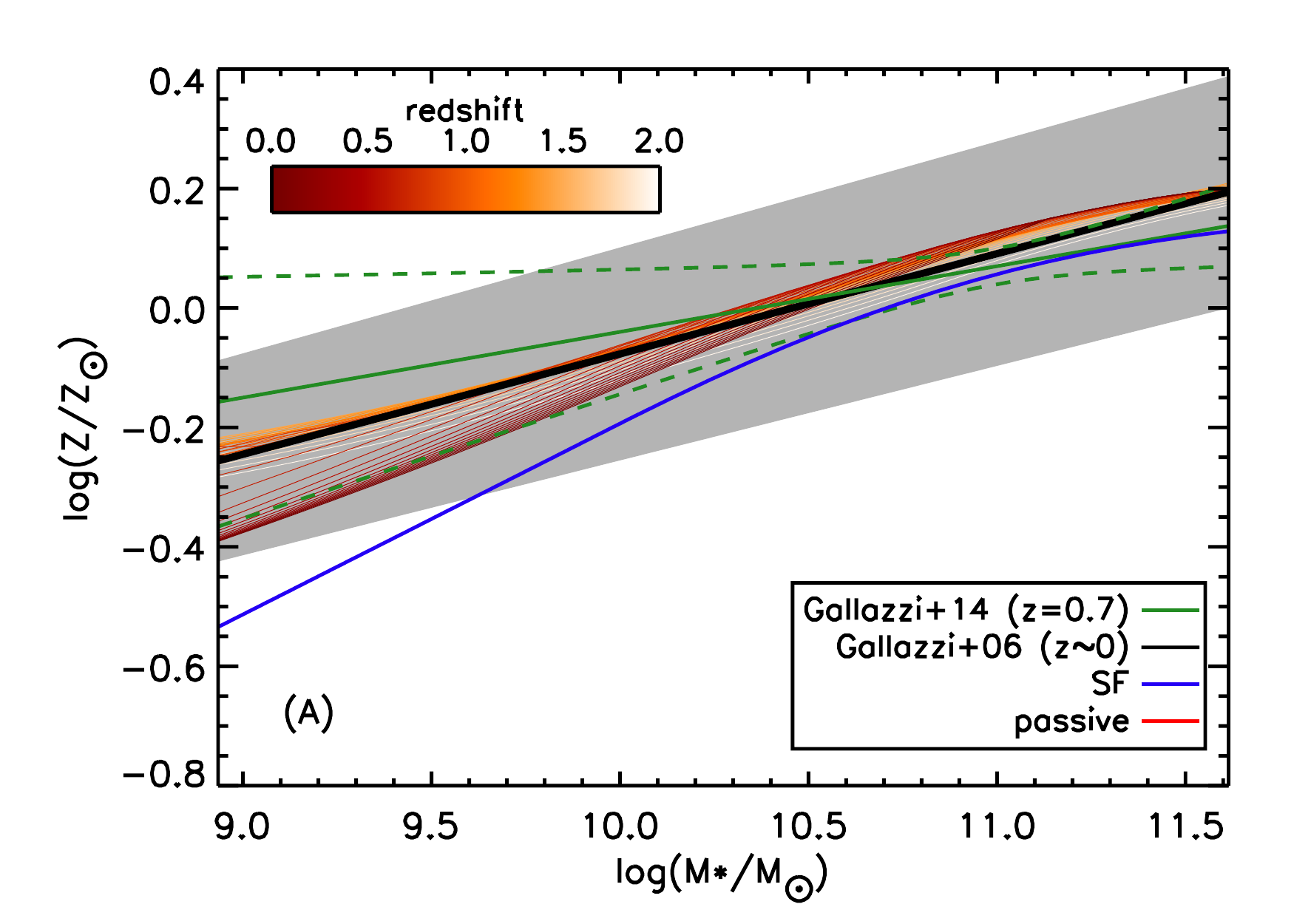}
\includegraphics[width=0.45\textwidth]{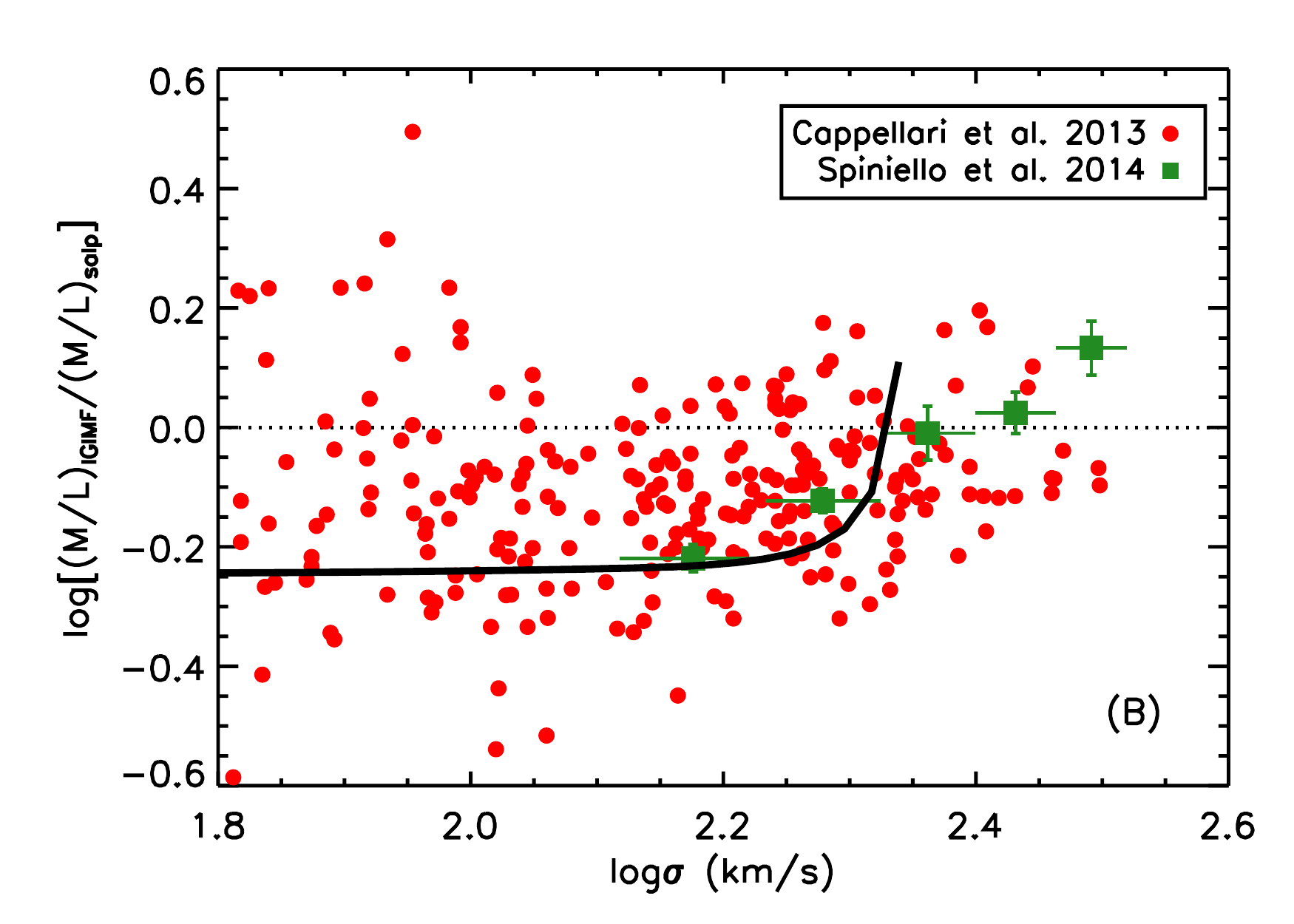}
\caption{\emph{(A):} stellar metallicity of ETGs, as a function of M$_{\star}$, predicted by our model (orange to red) 
and observed at $z\sim0$~and $z\sim0.7$ \citep{Gal06,Gal14} (black line with gray scatter, 
and green lines with dashed uncertainty envelope). For comparison, the stellar metallicity of SF galaxies at $z=0$ 
is shown in blue. 
\emph{(B):} M$_{\star}$ to bolometric light ratio (M/L) for $z=0$~passive galaxies (solid line), with respect 
to the Salpeter case (dotted line). The red dots and green squares show, respectively, spectroscopic observations 
of low-redshift ETGs from \citet{Cap13} and \citet{Spi14}. 
For consistency, we have converted M$_{\star}$~into velocity dispersions using Eq.~5 in \citet{Cap13}.
}
\label{fig:mmr}
\end{figure*}

In this model, passive galaxies follow the MS SFH until the start of quenching at $t_q$, after which we follow a 
simple closed box formalism, with initial gas masses and SFR efficiencies based on the 2-SFM parameterization 
\citep{Sar14}. The evolution of the SFR $\Psi$, gas content $M_{\text{g}}$, and metallicity $Z$~after $t_{\text{q}}$~are 
then given by the following equations, which we solve numerically in steps of $\Delta t=10$~Myr:

\begin{equation}\label{eq:quench}
\begin{aligned}
M_{\text{g}}(t>t_{\text{q}}) &= M_{\text{g}}(t_{\text{q}}) - \int_{t_{\text{q}}}^t\Psi(t')\mathrm(d)t'\\
\Psi(t>t_{\text{q}}) &= \epsilon(t-\Delta t) M_{\text{g}}(t-\Delta t)\\
Z_{g}(t>t_{\text{q}}) &= Z_{\text{g}}(t_{\text{q}}) - y\ln\left(\frac{M_{\text{g}}(t)}{M_{\text{g}}(t_{\text{q}})}\right)\\
\log\epsilon(t) &= (1-\beta_2)\log\Psi(t) - \alpha_2\\
\log M_{\text{g}}(t_{\text{q}}) &= \alpha_2 + \beta_2\log\Psi(t_{\text{q}})
\end{aligned}
\end{equation}

\noindent
where $\epsilon(t)$~and $M_{\text{g}}(t_{\text{q}})$~are, respectively, the SFR efficiency (SFE) at time $t$ and molecular 
gas mass at the start of quenching. As a result, the SFR will then decrease more or less rapidly depending on galaxy mass. 
We then consider that a galaxy has become \emph{passive} when its SFR is 1.5~dex below that of a same-mass galaxy on 
the MS. An example of a quenched SFH is shown in Fig.~\ref{fig:sfh}.
In our case, the values for the slope $\beta_2$~and intercept $\alpha_2$~of the $M_{\text{g}}$-SFR and SFE-SFR relations 
given in \citet{Sar14} would preclude the existence of passive galaxies with M$_{\star}$~below 
$\sim$5$\times10^{10}$~M$_{\odot}$~at $z\geq2$, in contradiction with observations \citep[e.g.,][]{Il13}. 
This is not entirely surprising, since environmental effects play a more significant role in the quenching of 
low-mass galaxies than high-mass ones even at high redshift \citep[][]{Stra13,Gob13,New14,Gob15} and can be expected 
to have different timescales than the simple case we are considering here. 
Letting both parameters vary, we find that values of $\alpha_2=8.8$~and $\beta_2=1$~reproduce better -within the 
constraints of the model- the observed ages and metallicities of ETGs 
\citep[see also Fig.~\ref{fig:mmr}]{Tho05,Gal06,Gal14}. 
These values imply a SFE lower than that of main sequence galaxies by a factor $\sim$3, consistent with the 
value reported at low redshift \citep{Mar13} and suggested by observation at $z\sim1.5$ \citep{Sar15}. 
We also verify that Eq.~\ref{eq:quench}~can reproduce reasonably well the observed stellar metallicity of passive 
galaxies at $z<1$~\citep{Gal06,Gal14}, as shown in Fig.~\ref{fig:mmr}. This can be achieved using a yield of 
$y\sim0.9$, regardless of the IMF (Sect.~\ref{imf}).\\

While we assume that SF galaxies constitute a homogeneous population that always follows MS evolution, passive 
galaxies clearly form a composite population of objects that have quenched at different times and thus have different 
SFHs. To include this fundamental aspect of the ETG population, we use observed galaxy mass functions (GMFs) from 
\citet[][at $z=0.2-2.5$]{Il13} and \citet[][at $z\sim0$]{Bal12}. The ETG population at $z$ is then given by the 
combination of the ETGs present at $z+\Delta z$, and the systems that quenched between $z+\Delta z$ and $z$, with their 
relative weights given by the value of the GMF, 

\begin{equation}\label{eq:gmf}
\begin{aligned}
x(M_{\star},z_j) &= \frac{\sum_{i=j}^{z_i\leq z_{\text{max}}} w_{i,j} x(M_{\star},z_i)}{\sum_i w_{i,j}}\\
w_{i,j} &= \frac{\Phi(M_{\star},z_i)-\Phi(M_{\star},z_{i-1})}{\Phi(M_{\star},z_j)}
\end{aligned} ,
\end{equation}

\noindent
where $\Phi$~is the GMF and $x$~a derived observable (such as luminosity, total M$_{\star}$, or number of stars or 
planets as in Eq.~\ref{eq:hab}) for a passive galaxy of mass $M_{\star}$~quenched at $z_i$. Here we choose 
$z_{\text{max}}=2$, making the approximation that all $z\sim2$~ETGs quenched at this redshift. 
For ease of computation, we evaluate the GMFs in discrete bins of redshift with width $\Delta z=0.05$.\\

We note that we do not consider galaxy mergers, except implicitly through the GMF and the SB fraction. The 
mass-weighted average SFH of two galaxies does not differ much from that of a galaxy of equivalent combined mass, and our 
assumed SFH can produce massive ETGs at high redshift without the need for mergers, with peak SFRs in their progenitors 
that are large (e.g., $\sim$600~M$_{\odot}$~yr$^{-1}$ for $\gtrsim5\times10^{11}$~M$_{\odot}$ ETG at $z=2$), but compatible 
with observations \citep[e.g.,][]{Fu13,Tan14,Sch15}. 
On the other hand, the metallicity of the final galaxy would be lower by up to $10-20$\% in the case of a merger, than 
that of a non-remnant galaxy. Furthermore, since higher mass galaxies are typically hosted in larger halos, with more 
surrounding substructure (i.e., satellites), galaxy mergers can be expected to play a more significant role in the case 
of large central galaxies \citep[e.g.,][]{Fel10,Car13}, and we can thus expect our estimates to match observables less 
well at large ($>10^{11}$~M$_{\odot}$) masses.

\subsubsection{\label{imf}Stellar initial mass function}

For ease of comparison with the rest of the literature, most studies assume a single universal stellar IMF, 
typically either the canonical \citet{Sal55} IMF or a ``bottom-light'' function 
\citep[these two IMFs have slightly different slopes and parameterization of the low-mass regime, but otherwise 
yield very similar M$_{\star}$~and SFRs]{Kro01,Cha03}. 
The choice of IMF typically only determines the scaling of derived quantities such as M$_{\star}$~or SFR, 
although it has a greater impact on the amount of metals produced and returned to the ISM.
On the other hand, recent studies have shown evidence for a varying IMF in massive elliptical galaxies 
\citep[e.g.,][]{vDk11,vDk12,Cap12,Cap13,Spi14}, implying that the IMF of their SF progenitors are not 
universal either. Furthermore, the IMFs mentioned above have all been derived by measuring the distribution of 
stars in localized regions of the MW and might therefore not be directly applicable to extragalactic environments.\\ 

Here we adopt the integrated galactic IMF \citep[IGIMF; ][]{KW03}, i.e., a convolution of the IMF of individual star 
clusters with a distribution of star cluster masses. The IMF of individual clusters is assumed to 
correspond to the \citet{Kro01} function below $m=1.3$~M$_{\odot}$, with a slope at $m>1.3$~M$_{\odot}$~and maximum stellar 
mass $m_{\text{max}}$~proportional to the embedded cluster mass \citep{Pfl07,Wei11}. The cluster mass function is, 
in turn, described by a power law with index $\beta=2$ with a maximum cluster mass depending on the galaxy's SFR 
\citep{Wei13}. Here we fix the minimum cluster mass at $50$~M$_{\odot}$. 
While clearly more complex, the IGIMF can reproduce some observables better than a fixed IMF 
\citep[see][and references therein]{Kro13}. In particular, since more massive galaxies experience higher peak SFRs than 
lower mass ones, the variation of the IGIMF slope with SFR naturally translates into a variation with final 
M$_{\star}$~when integrated over the galaxy's SFH. As an example, we show in Fig.~\ref{fig:mmr} the ratio of M$_{\star}$~to 
bolometric luminosity (M/L) of passive galaxies compared to those derived from observations of ETGs in the 
nearby universe. 
While not a tight fit, the IGIMF reproduces reasonably well the observed offset in M/L 
compared to the Salpeter case for low-mass galaxies. It also yields an increase in M/L at higher M$_{\star}$~(although 
the latter is steeper than observed, possibly due to our neglect of galaxy mergers). 
This occurs because the IGIMF is by design very similar to the Kroupa IMF at low SFRs, if slightly top-light due to 
the cutoff M$_{\star}$, but becomes top-heavy at SFR~$>100$~M$_{\odot}$~yr$^{-1}$. 
Since Eq.~\ref{eq:mssfh}, Eq.~\ref{eq:quench},and the FMR and GMFs, assume a Chabrier IMF, we  
evaluate these expressions using the appropriate IMF and convert all quantities to the IGIMF afterwards. 
For comparison, in Sec.~\ref{results}, we also give results with a Salpeter IMF. 
As noted in Sec.~\ref{sfh}, here we evaluate the IMF between 0.1 and 100~M$_{\odot}$.

\subsection{\label{hab}Habitable planets}

We assign a probability of hosting life-sustaining planets to the stars in each galaxy based on their age, metallicity and 
mass. What constitutes a habitable world, in the sense of a planetary body that can sustain life for some fraction of 
its lifetime, is somewhat conjectural as we are still limited by our incomplete understanding of the emergence and possible 
types of life. In principle, any environment allowing for liquid water and an energy source, whether internal or external, 
and stable over geological times could sustain life. 
These could include not only the usually considered icy moons of gas giants, but also more exotic environments such 
as free-floating planets \citep[e.g.,][]{Ste99,Lau00,AS11}. Furthermore, biochemistries different from our own could exist, 
in which case the range of conditions favorable to the emergence of life might be much greater than what we can currently 
envision, and all the more unconstrained.\\
 
For the purpose of this work, we limit ourselves to environments similar to our own, i.e., terrestrial planets 
around stars on the sMS, with a combination of atmospheric pressure and temperature allowing them to 
sustain liquid water on their surface. We adopt the definition of circumstellar HZ given by \citet{Kop13}, 
in this case the region bracketed by the ``moist greenhouse'' and ``maximum greenhouse''  limits(respectively, inner and outer). 
\citet{Kop13} define the HZ for effective temperatures between 2600 and 7200~K, corresponding to a mass interval 
of $\sim$0.1-1.5~M$_{\odot}$. We accordingly adopt a lower mass limit of 0.1~M$_{\odot}$ and make the ad hoc assumption that 
only stars with $\leq1.5$~M$_{\odot}$~can host habitable planets.  
Furthermore, we only consider stars older than $t_{\text{min}}=1$~Gyr as possible hosts for habitable planets, motivated by 
the presumably hostile surface conditions a planet would experience in its early eon, due to impacts from the large amount of 
debris left over from the protoplanetary disk and perturbed by planetary migration \citep[e.g.,][]{Gom05}.
In addition, the HZ shifts during a star's lifetime on the sMS as the latter brightens and its effective temperature changes. 
We take the effect of stellar aging into account by calculating the inner and outer HZ radii at each time step, using the 
fitting formulae of \citet{HPT00} to estimate accurate luminosities and effective temperatures.\\

The probability of a star hosting a terrestrial planet within its HZ is then given by the width of the HZ, the probability of 
planet formation, and the distribution of orbital periods. 
Despite recent advances, our knowledge of the distribution of low-mass or long-period exoplanets is still limited. In this 
paper, we mainly rely on the observed properties of short-period planets and use extrapolation when necessary. 
We assume that the distribution of the orbital periods $P$~of terrestrial planets follows a simple power law of the form 
$\mathrm{d}N\propto P(r,m)^{\beta_{\text{P}}}\mathrm{d}P$~($r$~and $m$~are, respectively, the radius of the orbit and 
mass of the host star) and set $\beta_{\text{P}}=-0.7$~in accordance with observations \citep[e.g.,][]{Cum08,Pet13,Bur15}. 
We also make the assumption that the fraction of stars with terrestrial planets, $f_{\text{T}}$, depends principally on the 
host star's metallicity rather than its mass. We consider two cases for this metallicity dependence:

\begin{equation}\label{eq:pter}
f_{\text{T}}(Z) = \mathcal{H}(Z-Z_{\text{min}}) 
\begin{cases}
f_{\text{T},0} (Z/Z_{\odot})^{\alpha_{\text{P}}} & \text{(Case 1)}\\
f_{\text{T},0} - f_{\text{HJ}} (Z/Z_{\odot})^{\alpha_{\text{P}}} & \text{(Case 2)}\\
\end{cases} .
\end{equation}

\noindent
Case 1 is a straightforward power law, based on the historical assumption that planet formation requires an abundance 
of heavy elements, and represents the simplest form of metallicity dependence for $f_{\text{T}}$. While Case 1 does not appear 
to fit very well the distribution of currently known \emph{terrestrial} planets, it still describes accurately the observed 
correlation between the occurrence of giant planets and the host star metallicity \citep[e.g.,][]{FV05,Gon14,GM14}. 
Because the value of the exponent is still somewhat uncertain, we choose a mid-range $\alpha_{\text{P}}=2$~for simplicity.  
On the other hand, the presence of non-giant planets does not seem to be strongly correlated with stellar metallicity 
\citep[e.g.,][]{Buc12,WF15,Buc15,Schu15}. Furthermore, the inward migration of a giant planet towards its host star would likely 
disrupt the formation of smaller planets or, if already present, their orbits. We can then simply assume that stars with 
short-period giants known as hot Jupiters cannot have terrestrial planets in their HZ 
\citep[but see, e.g.,][]{FN09,Bec15}. In Case 2 we therefore take a constant occurrence rate for terrestrial planets, 
but subtract from it the metallicity-dependent fraction of stars with hot Jupiters, $f_{\text{HJ}} (Z/Z_{\odot})^{\alpha_{\text{P}}}$, 
and therefore assume that $f_{\text{T}}$~is weakly \emph{anticorrelated} with metallicity \citep[see, e.g.,][]{Pra08,Zac16}. 
We take $f_{\text{HJ}}=0.012$~for solar-type stars \citep{Wri12} and, in both cases, set $f_{\text{T},0}=0.4$~at $Z=Z_{\odot}$~
following \citet{Pra08} and in accordance with the occurrence rate of planets with a radius of $<2R_{\oplus}$~and a period 
of $P=5-500$~days derived by \citet{Pet13} from the \emph{Kepler} sample. 
In both cases, we adopt a low-metallicity threshold of $0.1Z_{\odot}$~under which the probability of forming terrestrial 
planets is zero \citep[e.g.,][]{FV05,Buc12}.
The fraction of stars of a given mass, age, and metallicity with habitable planets can then be expressed as 

\begin{equation}\label{eq:habweight}
w_{\text{h}}(m,Z,t) = f_{\text{T}}(Z) \mathcal{H}(1.5-m) C \int_{r_{\text{h,i}}(m,Z,t)}^{r_{\text{h,o}}(m,Z,t)} 
P(r,m)^{\beta_{\text{P}}} \frac{\mathrm{d}P}{\mathrm{d}r} \mathrm{d}r ,
\end{equation}

\noindent
where $C$~is a normalization constant so that $C\int N\frac{\mathrm{d}P}{\mathrm{d}r}\mathrm{d}r=1$~over the whole range of 
periods considered. Here we conservatively take a lower and upper limit of 3~days and 300~years, respectively.

\subsection{\label{snagn}Supernovae and AGN}

The habitable stellar fraction given in Eq.~\ref{eq:habweight} can be considered an upper limit, as energetic 
phenomena such as supernovae (SN) and active galactic nuclei (AGN) can sterilize the surface of planets if they 
happen close enough. 

\subsubsection{\label{snr}Supernovae}

The majority of supernova events falling into either the type Ia or type II category 
\citep[e.g.,][]{Hak14,Cpp15}, we only consider these two classes (hereafter, SNIa and SNII). 
Both the SNII and SNIa rate depend on the IMF through the fraction of stars with, respectively, 
$m\geq8$~M$_{\odot}$~and $m<8$~M$_{\odot}$. The SNII rate is directly proportional to the 
\emph{instant} SFR, the lifetime of a $m\geq8$~M$_{\odot}$~star being shorter than the typical timescale 
normally considered for star formation in galaxies ($\sim$100~Myr). On the other hand, SNIa are significantly 
delayed by the longer time spent on the sMS by their progenitors and the additional time elapsed between the 
formation of the white dwarf (WD) and its detonation. The rates of the two SN types can thus be expressed as:

\begin{equation}\label{eq:SNrate}
\begin{aligned}
\text{SNR}_{\text{II}}(t) &= \Psi(t)\int_{8 \text{M}_{\odot}}^{m_{\text{max}}} m\phi(m,t)\mathrm{d}m\\
\text{SNR}_{\text{Ia}}(t) &= \eta_{\text{WD}} \int_{m_{\text{min}}}^{8 \text{M}_{\odot}} m 
\int_{\tau_{\text{Ia}}+t_{\text{MS}}(8)}^t \Psi(\tau) \phi(m,\tau) \mathrm{d}t' \mathrm{d}m
\end{aligned} ,
\end{equation}

\noindent
where $\eta_{\text{WD}}=0.01$ is the WD conversion rate \citep{Pri08} and 
$\tau=t'-t_{\text{MS}}(m)-\tau_{\text{Ia}}$, with $t_{\text{MS}}(m)$~and $\tau_{\text{Ia}}=500$~Myr being, 
respectively, the main sequence lifetime of a progenitor of mass $m$~and the average delay time between stellar 
death and the detonation of the WD \citep{Ras09}.\\

The impact of SN on galaxy habitability depends principally on two parameters: the distance $r_{\text{SN}}$~at which 
a SN can sterilize a planet and/or render it uninhabitable (hereafter the lethal radius) and the time required 
by a planet to recover from the effects of the SN, $t_{\text{rec}}$. 
There is circumstantial evidence that SN events might be implicated in some of the mass extinctions of species that 
have happened on Earth throughout its history \citep[e.g.,][]{ES93,Det98,Mel09,Th15}, but the full range of SN effects 
on life-bearing worlds has, to our knowledge, not been extensively probed. 
The emphasis is usually put on hard radiation (e.g., gamma rays) owing to its disruptive effect on atmospheric ozone, 
which leads to an increase in biologically hazardous UV irradiance. For example, \citet{Gow11} and \citet{For15} use 
the ozone-depletion estimates of \citet{Geh03} to set the lethal radius at 8~pc. However, the expected rate of SN 
within 8~pc \citep[see][]{Geh03} suggests that the Earth might have experienced several during the last few Gyr 
while still ultimately retaining its capability to support life. 
Such an event would also only affect the unprotected areas of a planet: photosynthesis-independent benthic and 
lithoautotrophic organisms would be scarcely affected by hard radiation on the surface and the approximately tenfold 
decrease (in the case of Earth) in ocean water absorbance between UVB wavelengths and the first peak of chlorophyll 
absorbance 
\citep[$\sim$300~nm and $\sim$450~nm, respectively; see][]{Shi88} suggests that even shallow marine ecosystems would 
be somewhat more resistant than surface ones.
We can then surmise that, at 8~pc, $t_{\text{rec}}$~is of the order of the recovery time of biodiversity after mass 
extinctions as inferred from the fossil record, i.e., small with respect to the timescales relevant to galaxies. 
In the $\{r_{\text{SN}}=8\text{pc},t_{\text{rec}}\sim0\}$~case, the effect of SN on habitability averaged over a 
whole galaxy is then likely negligible for all but the most active star-forming galaxies.\\

Instead, we focus here on thermal effects, i.e., how far a SN can occur from a planet and still push it out of the HZ 
of its host star. Since this distance depends on the position of a planet within the HZ and on the 
width of the HZ, we define the lethal radius as the distance at which 50\% of planets normally within 
the HZ of their host star find themselves out of it for the duration of the event. 
We consider two different sources of heating: the initial bolometric radiation emitted by the SN over $\sim$2~months 
and the blastwave arriving some time later. In the first case, we assume a radiative output of $10^{49}$~erg for SNII 
and $3\times10^{49}$~erg for SNIa \citep{Sca14}. In the second, we consider $\sim$10~M$_{\odot}$ of 
ejecta for SNII \citep[e.g.,][]{Woo02} and 1.4~M$_{\odot}$~for SNIa, adiabatically expanding with an 
initial kinetic energy of $10^{51}$~erg. We assume that the blastwave propagates in a surrounding ISM of pure hydrogen 
with an average density of 1~atom/cm$^3$~\citep[see, e.g.,][]{Dra11}.
We further posit that the ejected shell can only interact with a planet as long as its pressure is larger than that 
of the wind of the host star, for which we assume a constant density of 10 atoms/cm$^3$~and a speed of $500$~km/s, 
and that the entire kinetic energy of the gas will be converted into heating the planet's atmosphere. 
Because the bolometric radiation is mostly released in the days following the detonation, but the ejected material 
expands much more slowly, these two effects are considered independently. 
For our choice of $\mathrm{d}N/\mathrm{d}P$, we find a lethal radius of $r_{\text{Ia}}=0.3$~pc for SNIa and 
$r_{\text{II}}=0.5$~pc for SNII. 
We make the additional assumption that such an excursion from the HZ will sterilize a planet and permanently alter 
its surface conditions, rendering it inhospitable, and thus set the recovery time $t_{\text{rec}}$~to a value greater 
than the Hubble time $t_H$. 
On the other hand, feedback processes in the planet's atmosphere might, depending on its composition, compensate 
for the increased energy input. However, ascertaining the actual impact of the SN would then require complex 
modeling that is beyond the scope of this paper.\\

We then estimate a fractional irradiated volume to use as a correction term to the time-varying habitable 
fraction. We approximate passive galaxies as spheres and SF galaxies as disks, both with constant 
stellar density and redshift-dependent radius set by the observed mass-size relation \citep{vdW14}. 
We assume that the disks have a thickness of $h_{\text{d}}=0.25\times r_{\text{e}}$~at high redshift, where 
$r_{\text{e}}$~is the effective radius of the galaxy, and transition to a thin disk with 
$h_{\text{d}}=0.15\times r_{\text{e}}$~\citep[based on the structure of the MW disk;][]{Bov12} after the peak 
of their SFH \citep{Leh14}. The fractional irradiated volumes for passive (P) and SF galaxies are then 

\begin{equation}\label{eq:virr}
V_{\text{irr}}(t) = \mathcal{H}(t_{\text{rec}}-t)
\begin{cases}
\text{SNR}_{\text{Ia}}(t) \frac{r_{\text{Ia}}^3}{r_{\text{e}}^3(t)} & \text{(P)}\\
\frac{4}{3h_d(t)r_{\text{e}}^2(t)}(\text{SNR}_{\text{II}}(t) r_{\text{II}}^3 + \text{SNR}_{\text{Ia}}(t) 
r_{\text{Ia}}^3) & \text{(SF)}\\
\end{cases} .
\end{equation}

\noindent
Since we do not include the growth of passive bulges in disk galaxies, we simply assume that the volume of 
ETGs is given by the disk case during their SF phase and the passive one after quenching, with a simple 
linear transition between both during quenching.
We assume internal homogeneity for simplicity. This is however a clear limitation, especially in the 
case of SF galaxies, as we can expect SNII and SNIa to preferentially happen in comparatively dense 
star clusters \citep[e.g.][]{Sha02,Mao10}, which would increase SN lethality. In more extreme environments, 
such as massive star clusters or very compact galaxies, nearby main sequence stars might contribute enough 
to the ambient radiation field to adversely affect the HZ \citep{Thm13,Ad15}.\\ 

\subsubsection{\label{AGN}AGN}

We estimate a lethal radius $r_{\text{AGN}}$~resulting from the activity of a galaxy's central supermassive black 
hole (SMBH) in a similar manner. Although this activity is expected to be sporadic, we assume it to be frequent 
enough to define an exclusion region where planets cannot stay habitable over the long term. In this case, only 
thermal radiative effects are considered. We assume that the SMBH radiates at Eddington luminosity and use the 
local relation of \citet{RV15} for ellipticals and bulges to derive the SMBH mass from a galaxy's M$_{\star}$. 
The distance from the SMBH at which 50\% of planets fall out of the HZ of their host star depends on the AGN 
luminosity, thus M$_{\star}$~through the relation 
$\log r_{\text{AGN}} (\text{kpc})=-6.1+0.7\times\log M_{\star} (\text{M}_{\odot})$. We then add the spherical 
volume defined by this radius to the value of $V_{\text{irr}}$~given in Eq.~\ref{eq:virr}.

\subsection{\label{galhab}Galaxy habitability}

Combining all the elements described in the previous sections, we define the habitability $h_{\text{G}}$ of a galaxy 
at redshift $z$~as the ratio of the number of main sequence stars of mass $\leq1.5$~M$_{\odot}$, age 
$\leq t_{\text{min}}=1$~Gyr, and with planets in their HZ, to the total number of stars present in the galaxy at $z$. 
In a slightly shortened notation 

\begin{equation}\label{eq:hab}
h_{\text{G}} = \frac{
\int_0^{t_{\text{z}}-t_{\text{min}}} (1-V_{\text{irr}}(t)) \Psi(t) \int_{m_{\text{min}}}^{1.5} 
m\phi(m,t)\mathcal{H}(t_{\text{MS}}(m)-t) w_{\text{h}} \mathrm{d}m \mathrm{d}t
}{
\int_0^{t_{\text{z}}} \Psi(t) \int_{m_{\text{min}}}^{m_{\text{max}}} m\phi(m,t)\mathcal{H}(t_{\text{MS}}(m)-t) 
\mathrm{d}m \mathrm{d}t 
}
\end{equation}

\noindent
for SF galaxies. The habitability $h_{\text{G}}$~is here a function of the time $t_{\text{z}}$~after the onset of star formation 
and, through the SFH $\Psi$, the M$_{\star}$~of the galaxy at $z$. In the case of passive galaxies, both sides of 
the fraction are a combination of different age populations as described in Eq.~\ref{eq:gmf}.

\section{\label{results}Results}

\begin{figure*}
\centering
\includegraphics[width=0.45\textwidth]{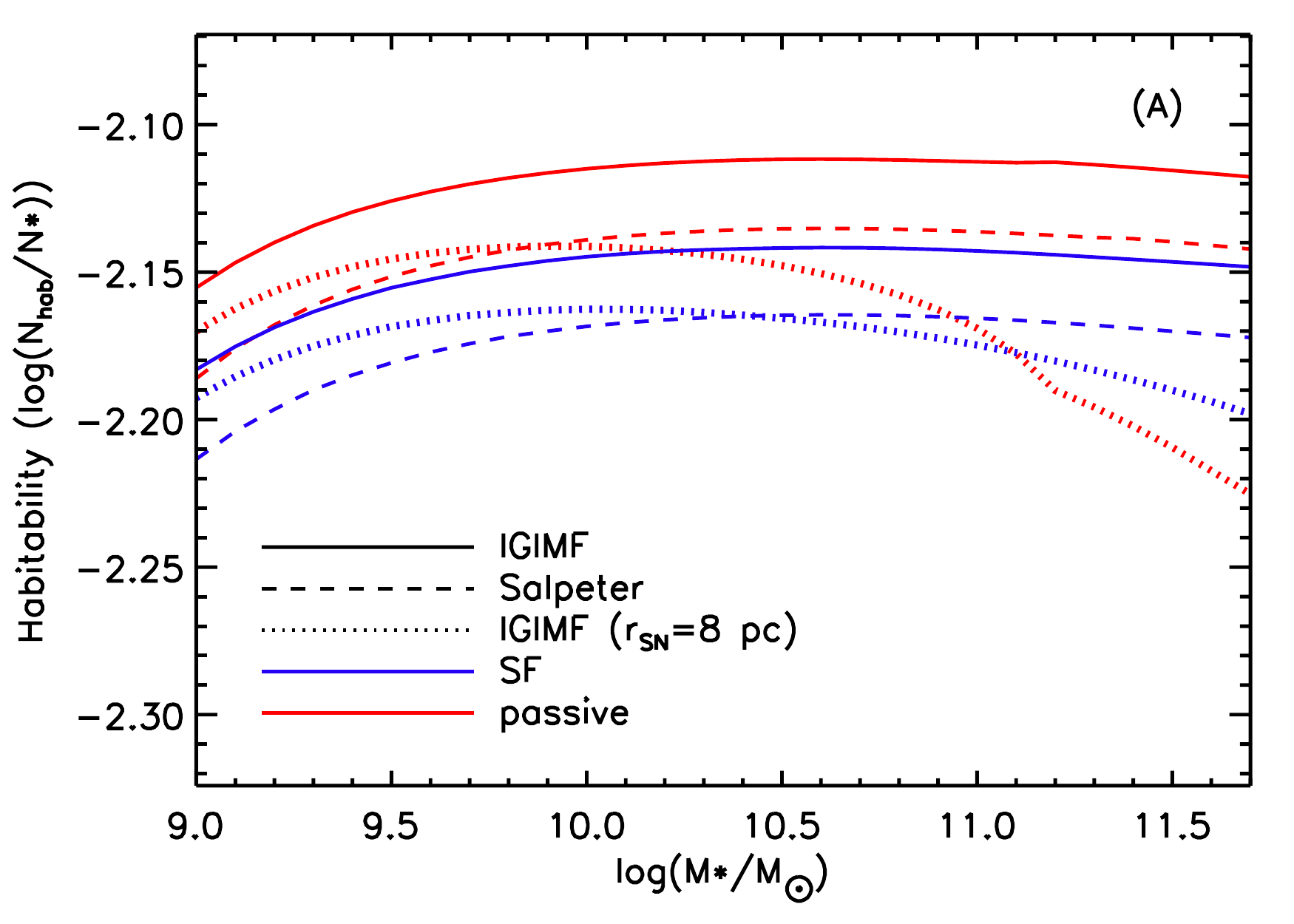}
\includegraphics[width=0.45\textwidth]{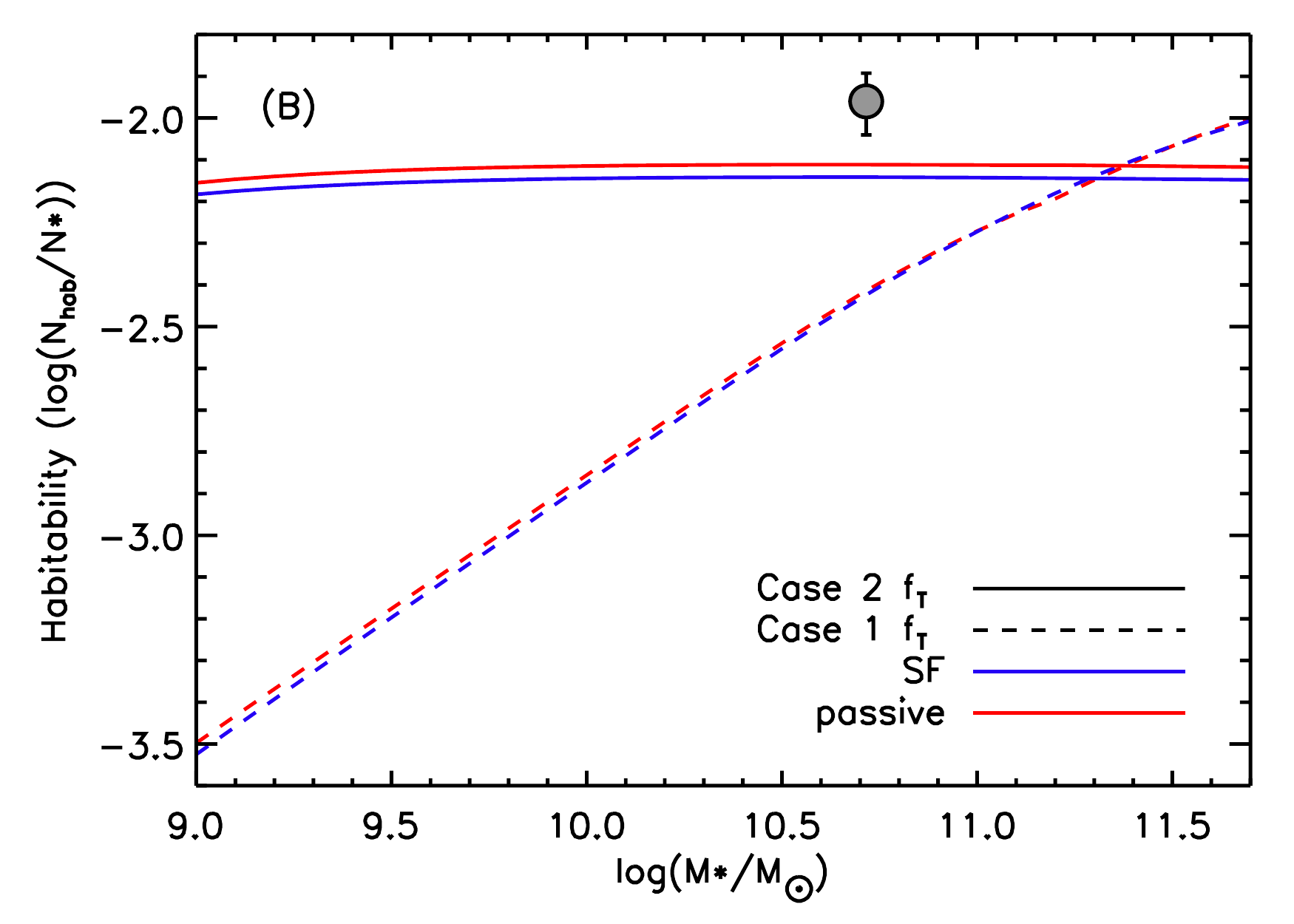}
\includegraphics[width=0.45\textwidth]{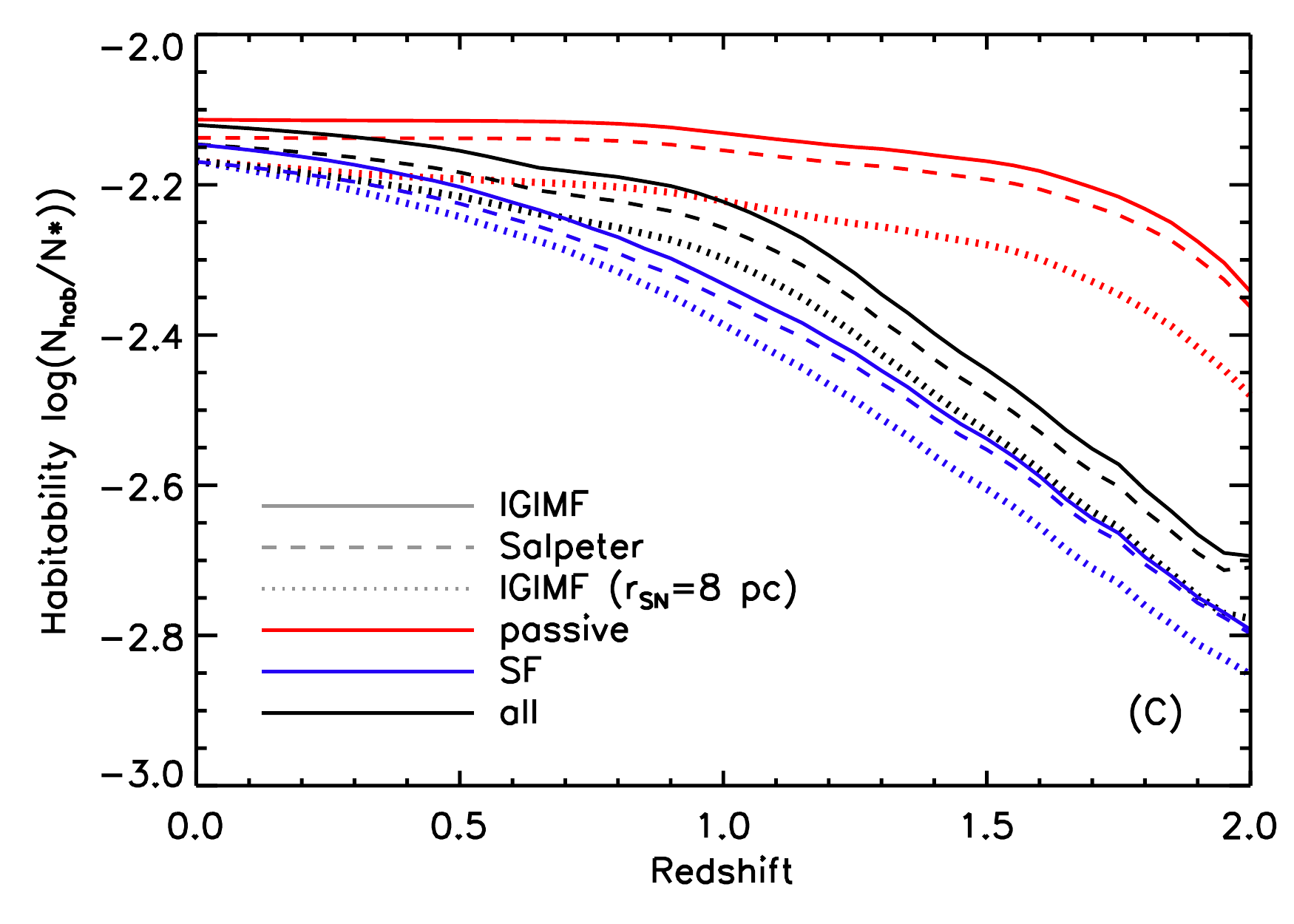}
\includegraphics[width=0.45\textwidth]{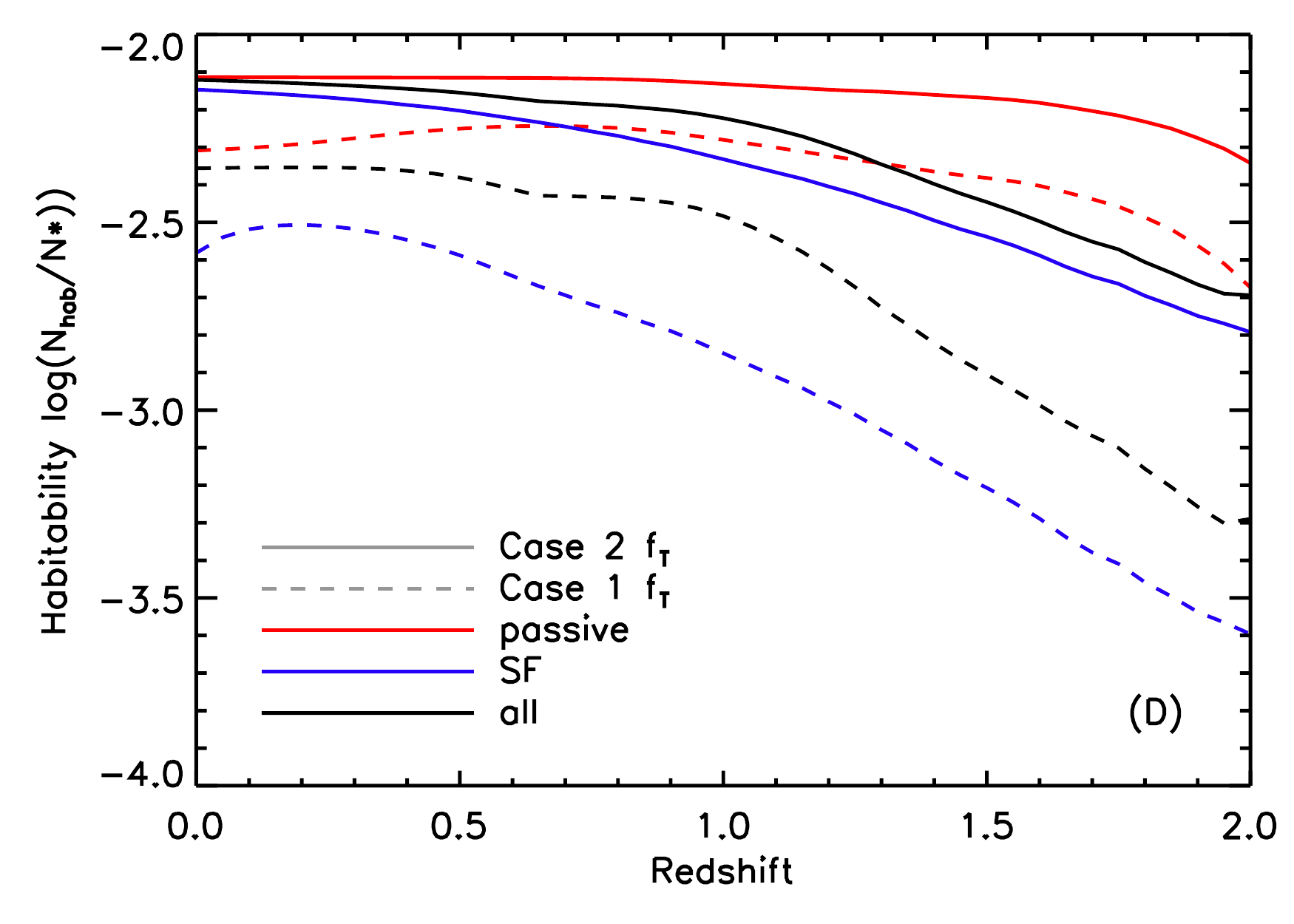}
\caption{\emph{(A):} Galaxy habitability at $z=0$~as a function of mass and IMF (IGIMF and Salpeter; 
solid and dashed lines, respectively), for SF (blue) and passive (red) galaxies and a 
metallicity-anticorrelated (Case 2) $f_{\text{T}}$. To illustrate the effect of SN on the mass-dependency of 
$h_{\text{G}}$, the dotted lines show the IGIMF case with the lethal radius of \citet{Geh03} 
$r_{\text{II}}=r_{\text{Ia}}=8$~pc and $t_{\text{rec}}>t_H$. 
\emph{(B):} Galaxy habitability at $z=0$~of SF (blue) and passive (red) galaxies, using the 
IGIMF and for the two different cases of terrestrial planet incidence (Case 1 and 2; dashed and solid 
lines, respectively). The filled gray circle shows the ratio of stars with terrestrial planet candidates 
in the habitable zone to the total number of stars with at least one planet candidate, taken from the 
NASA Exoplanet Archive using the same criteria (star mass, illuminance, planetary radius) as described in 
Sect.~\ref{hab}. The error bar assumes Poisson uncertainties.
\emph{(C):} Galaxy habitability integrated over the range of galaxy masses, as a function of 
redshift and IMF (IGIMF and Salpeter; solid and dashed lines, respectively), for SF (blue) 
and passive (red) galaxies. As in (A), this panel assumes a Case 2 $f_{\text{T}}$. The dotted 
lines show the evolution of habitability in the IGIMF case if we use the larger (8~pc) SN lethal 
radius of \citet{Geh03}.
\emph{(D):} As in (C), evolution of galaxy habitability with redshift as a function of the 
metallicity dependence of $f_{\text{T}}$. In both panels, the black lines show the evolution of habitability 
averaged over the whole galaxy population (SF and passives).
}
\label{fig:hmz}
\end{figure*}

Figure~\ref{fig:hmz} shows the galaxy habitability $h_{\text{G}}$, computed in the mass range $10^9<M<3\times10^{11}$~M$_{\odot}$, 
as a function of M$_{\star}$~at $z=0$ and as a function of redshift. We consider several combinations of 
parameters: the IGIMF with Case 1 and 2 metallicity dependence as defined in Eq.~\ref{eq:pter} (panels B and D), 
and Case 2 only with either an IGIMF or Salpeter IMF (panels A and C). 
For Case 2 at $z=0$, between 0.65\% and 0.8\% of all stars host potentially habitable planets and $h_{\text{G}}$ is only weakly 
dependent on M$_{\star}$, with a maximum around $4\times10^{10}$~M$_{\odot}$ 
\cite[consistent with][]{Zac16}. The habitability of ETGs is $\sim$0.03~dex higher than for disks of similar mass as a 
consequence of the 1~Gyr delay time before a planet becomes habitable. This reflects the typical fraction of stars formed 
in disks since that look-back time.
The main factors determining the overall normalization and behavior of $h_{\text{G}}$~are, in order of importance:\\
$\bullet$~{\bf metallicity}: lower mass galaxies also have lower metallicity and therefore a higher fraction of 
stars below the threshold for terrestrial planets, while for higher mass galaxies the decrease in $h_{\text{G}}$~stems from 
$f_{\text{T}}$~being inversely proportional to metallicity. 
In contrast, the habitability for Case 1 increases monotonically with M$_{\star}$~and is comparable in behavior to 
the model of \citet{Day15}. 
The $M_{\star}-Z_{\star}$~relation is well defined for both SF and passive galaxies, with an intrinsic scatter whose 
value is not precisely known, although it is believed to be small. Using a value of 0.1~dex close to the \emph{observed} 
scatter would add a dispersion of 0.03~dex and 0.008~dex at the low- and high-mass end of Fig.~\ref{fig:hmz}, respectively, 
for Case 2, and a constant dispersion of 0.2~dex for Case 1.\\
$\bullet$~{\bf IMF}: while metallicity drives the mass dependence of $h_G$, the IMF determines its average 
value because the number of habitable planets in a stellar population depends on its fraction of low-mass stars. For example, 
using a Salpeter IMF decreases the habitability by $\sim$0.02~dex, owing to the lower fraction of 
$\lesssim1$~M$_{\odot}$~stars compared to the IGIMF, while the variation of $h_{\text{G}}$~with M$_{\star}$~is similar 
in both cases. The negative slope of the IMF in the range $0.1-100$~M$_{\odot}$~implies that most terrestrial planets will be 
hosted by subsolar mass stars, which have tighter HZs than solar mass stars. As a consequence, $h_{\text{G}}$~is rather sensitive 
to the slope of the distribution of orbital periods and an error of 0.1 in $\beta_{\text{P}}$~\citep{Cum08} would 
offset $h_{\text{G}}$~by 0.2~dex, making this parameter as important as $f_{\text{T},0}$~in determining its normalization.\\
$\bullet$~{\bf supernovae}: since our model assumes a small lethal radius, the effect of SN is very weak. 
It is slightly stronger at higher masses and for ETGs, which have SFHs that peak higher than disks of identical mass 
and at earlier times, when galaxies are denser. On the other hand, while we have ruled out large lethal radii in our 
model, the 8~pc value of \citet{Geh03} might still be relevant if we were to consider a more restrictive criterion for 
planetary habitability (see Sec.~\ref{civ}). As shown in Fig.~\ref{fig:hmz}, the impact of SN would in this case lead to 
significant differences in the $h_{\text{G}}$~of passive and SF galaxies at high masses. 
Active galactic nuclei behave in a similar fashion, fractionally increasing the irradiated volume in high-mass galaxies. 
The larger irradiated volume and fraction of stars below the metallicity threshold make these two effects stronger at 
higher redshift as well.\\

Other sources of uncertainty may include the GMF and the various parameters describing the distribution of terrestrial 
planets in Sec.~\ref{hab}. They are, however, inconsequential compared to the ones discussed above: the uncertainty generated 
on the $h_{\text{G}}$~of passive galaxies and its evolution by the errors on the GMF parameters is negligible, while the 
errors on $f_{\text{HJ}}$~and $\alpha_{\text{P}}$~\citep[0.038 and $\sim$1;][]{Wri12,FV05,Gon14} induce an uncertainty 
of at most 0.005~dex and 0.003~dex, respectively, across the mass range we consider. 
The uncertainties associated with planetary distribution functions arise mostly from instrumental limitations and small 
sample sizes. We can thus expect them to be mitigated in the near future as complementary surveys will extend our coverage 
of parameter space and increase existing samples by orders of magnitude 
\citep[e.g., \emph{Gaia} for long-period Jovians and PLATO for intermediate-period terrestrials;][]{Per14,Rau14}.\\

We compare our estimates with observations from the \emph{Kepler} mission using the NASA Exoplanet 
Archive\footnote{http://exoplanetarchive.ipac.caltech.edu}. If the \emph{Kepler} sample is mostly unbiased 
and terrestrial planets have the same distribution of orbital inclinations as giant ones, the ratio of stars with 
planet candidates within the HZ to the total number of stars with planet candidates should represent an estimate 
of $h_{\text{G}}$~(if all stars host at least one planet) or an upper limit to it.
We first select all stars observed by \emph{Kepler} with at least one confirmed or candidate planet and 
$T_{\text{eff}}\leq7200$~K, in accordance with Sec.~\ref{hab}. After taking multiplanetary systems into account, 
we find 3570 such objects in the database. This corresponds to 3924 stars, after correcting for the fraction of 
stars with $<1.5$~M$_{\odot}$~to the total and assuming a \citet{Kro01} IMF for simplicity.
We repeat the procedure adding conditions for the planet's radius ($<2R_{\oplus}$) and insolation (between 0.22 
and 1.13 times that of Earth) consistent with Sec.~\ref{hab}, and find 44 candidates. This corresponds to a ratio 
of $\sim$1\%, slightly higher but broadly consistent with our Case 2 estimate 
\citep[assuming a mass of $5.2\times10^{10}$~M$_{\odot}$~for the MW disk;][]{LN15}.
This is not unexpected, since our model uses parameters derived from the \emph{Kepler} sample. It would however 
tend to validate the use of Case 2 and our assumptions for galaxy evolution, as well as our chosen combination of 
$f_{T,0}$~and $\beta_{\text{P}}$. On the other hand, the Case 1 prediction lies significantly below the observed 
ratio, and can be reconciled only if $f_{\text{T},0}=1$, i.e., if all stars of solar metallicity or greater host a 
terrestrial planet, in conflict with observations.\\

\begin{figure}
\centering
\includegraphics[width=0.45\textwidth]{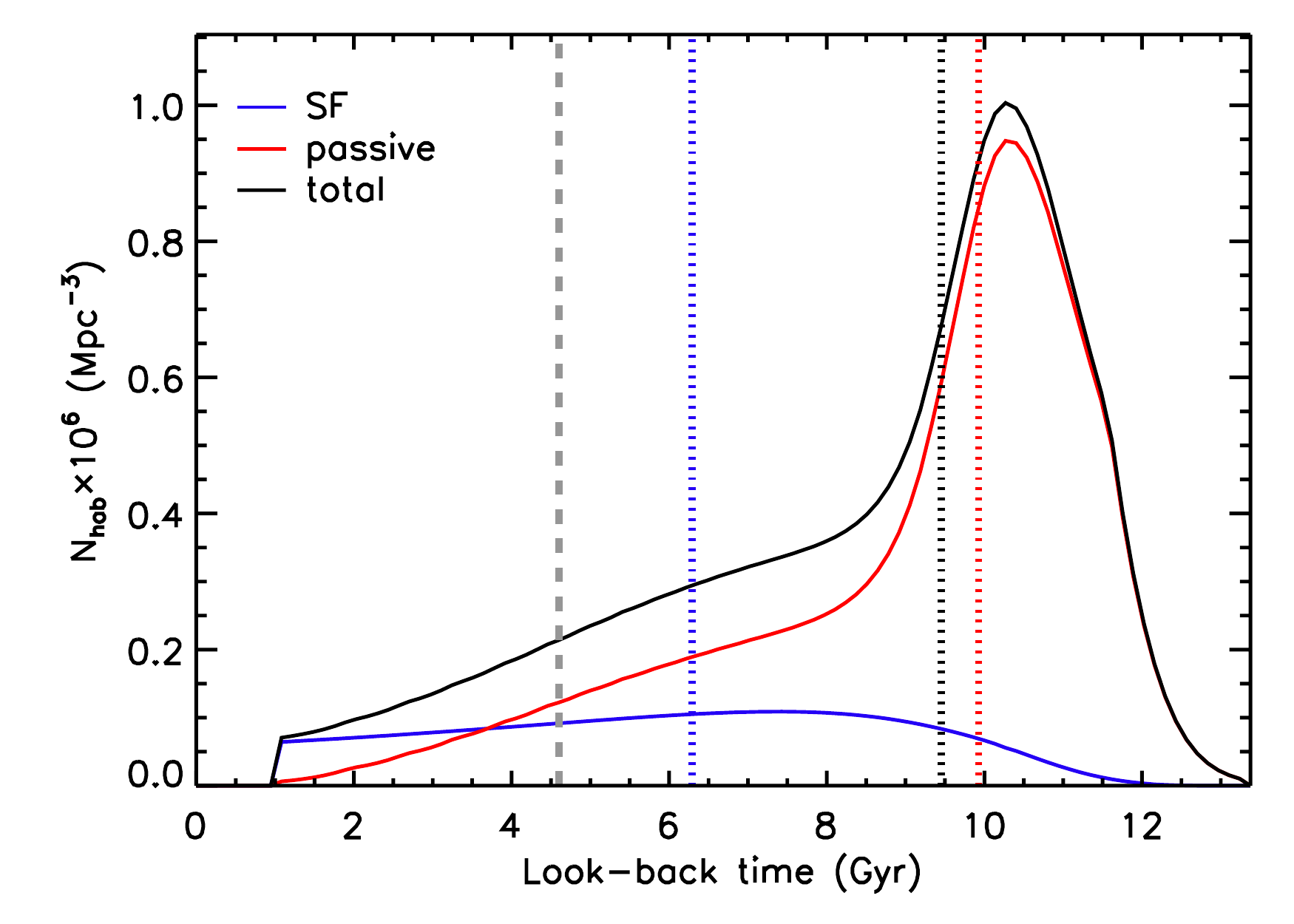}
\caption{Distribution of the ages of habitable planets per unit volume at $z=0$, for SF 
(blue) and passive (red) galaxies and with the total distribution shown as a black curve. The dotted 
lines show the respective median ages, while the dashed gray line marks the formation time of 
the solar system.
}
\label{fig:ages}
\end{figure}

In panels C and D of Fig.~\ref{fig:hmz} we show the redshift dependence of $h_{\text{G}}$, integrated over the mass range 
$10^9-3\times10^{11}$~M$_{\odot}$. The evolution of the integrated galactic habitability is monotonic with time 
for both SF and passive galaxies, and in both cases can be broadly divided into an early linear rise 
followed by a plateau in which $h_{\text{G}}$~increases by no more than $0.1$~dex, with the transition happening at 
$z>1.5$~for passive galaxies and $z<0.5$~for SF galaxies. We note that for Case 1 the integrated habitability is 
very sensitive to the number ratio of low-mass, low-metallicity (and hence low-$h_{\text{G}}$) galaxies to high-mass, 
high-metallicity (and high-$h_{\text{G}}$) ones. Therefore, the apparent decrease in habitability at 
$z<0.2$~for SF galaxies shown in Fig.~\ref{fig:hmz}, panel D, is likely due to our use of different GMFs for 
$z\sim0$~and $z>0.2$~populations. On the other hand, extending our SFHs to future epochs, and assuming that 
the FMR holds, we find that the habitability of SF galaxies for Case 2 will asymptotically approach that of 
passive galaxies, increasing by just $\sim$0.02~dex over the next 5~Gyr. 
For the total galaxy population, the shift occurs at $z\sim1$, i.e., after the peak of SFR in galaxies 
\citep[and thus of cosmic SFR; e.g.,][and references therein]{Mad14}. Since the stellar density does not appear 
to have changed significantly since this epoch as well, this implies that the total number of  habitable planets 
in a given cosmological volume has remained roughly constant since $z\sim1$, i.e., in the last 7-8~Gyr. 
In the local universe, most of the stellar mass is held in large passive galaxies that formed at $z>1$. 
This suggests that most of the habitable planets in the present epoch should belong to stars older than our own. 
Indeed, as shown in Fig.~\ref{fig:ages}, the median age of stars hosting habitable planets at $z=0$~is $\sim$9.5~Gyr, 
5~Gyr older than our own Sun and close to the estimate of \citet{Beh15}. 
On the other hand, if we only consider SF galaxies the median age drops to $\sim$6~Gyr, or $\sim$1.5~Gyr older 
than the Sun, which formed after $\sim$70\% of stars in disks that host habitable planets at $z=0$, 
in agreement with the earlier estimate of \citet{Lin01}. 
Unlike the work presented here, the studies referenced above did not consider stellar death. However, the shape 
of the IMF implies that most planets orbit long-lived subsolar mass stars, greatly mitigating its effect. On the 
other hand, the impact of stellar death would be more important when considering timescales longer than the Hubble 
time or non-standard IMFs. 
Finally, we note that we might be overestimating the number of presently habitable planets formed near the 
peak of SF in passive galaxies ($z\sim2$, or $\sim$10~Gyr ago) as a result of the simplistic assumptions on galaxy 
structure that our model uses, as stated in Sec.~\ref{snr}.

\subsection{\label{civ}On the plurality of Earths}

In the previous sections we used a rather broad definition of planetary habitability, which likely includes 
environments that are at best marginal to life (e.g., planets that spend only a short amount of time in the 
HZ). However, we as a culture seem to be most fascinated by the possibility of existence of other worlds and 
\emph{intelligences} similar to our own. The latter in particular has become a recurrent part of both our folklore 
and scientific discourse \citep[e.g.,][]{Dys60,Hew68,Bow82,Wri14}. Complex surface ecosystems such as the ones 
present on Earth today, let alone intelligent species, require time to arise by random evolution and are more 
sensitive to outside catastrophes (e.g., SN) than life in general. If we simply assume that the emergence 
of complex and/or intelligent life on a habitable planet is not contingent upon the type and 
metallicity of its host star, the frequency of Earth-analog ``garden worlds'' and that of intelligent life 
(hereafter civilizations) should then be directly proportional to the number of habitable planets, with 
two modifications: First, we increase $t_{\text{min}}$, with the added condition that the planets stay within the 
(shifting) HZ for a period at least equal to $t_{\text{min}}$ (e.g., the rise of our own species corresponds to 
$t_{\text{min}}=4.5$~Gyr). This accounts for the delay between the appearance of life and that of complex organisms; 
for the latter to happen, conditions must remain favorable throughout. 
Second, in addition to the SN lethal radius described above, we include the 8~pc irradiation radius of 
\citet{Geh03} with a recovery time of $t_{\text{rec}}=50$~Myr to account for SN-induced mass extinctions. 

\begin{figure}
\centering
\includegraphics[width=0.45\textwidth]{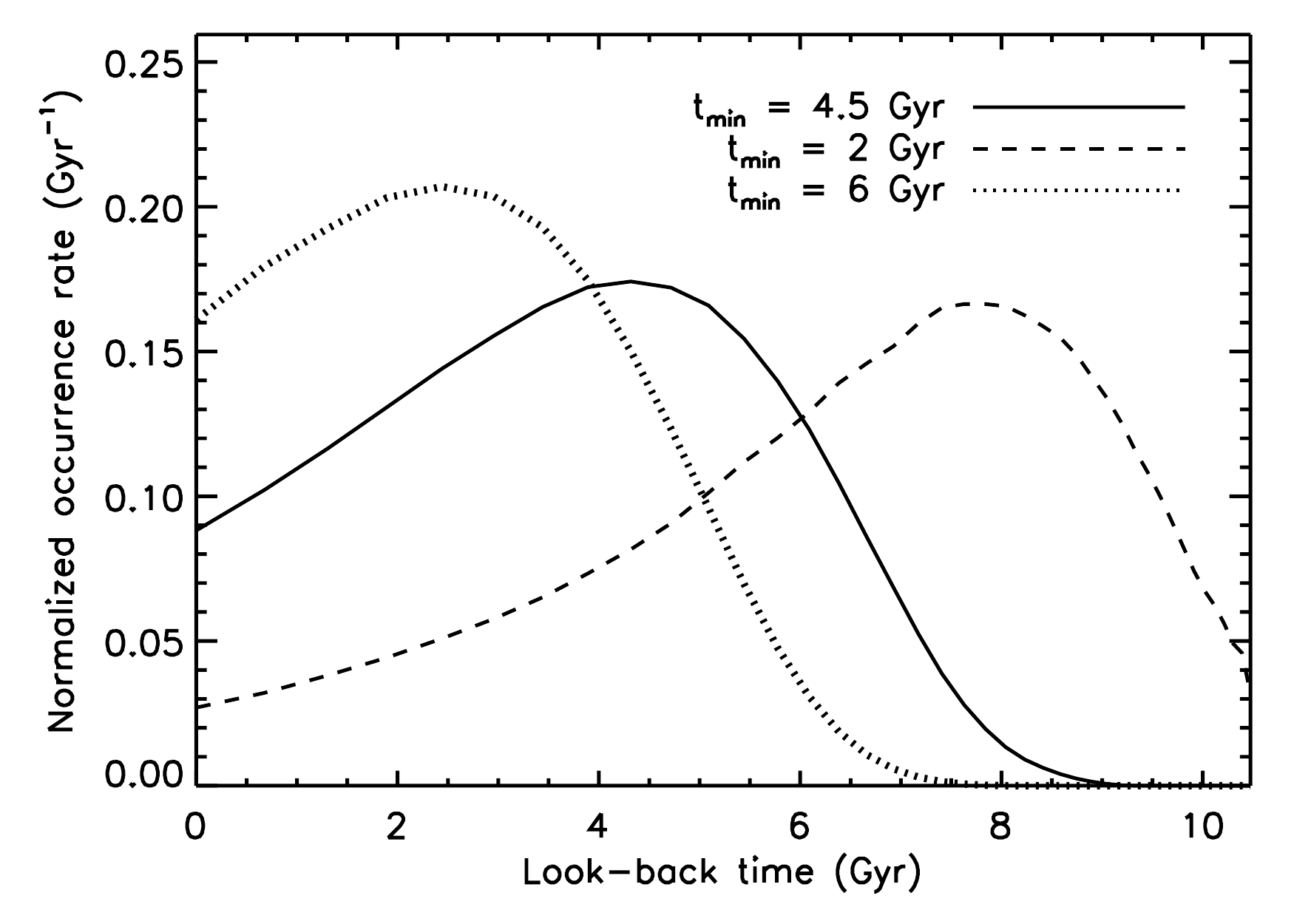}
\caption{Occurrence rate of civilizations in a MW-type disk, normalized to 1 at $z=0$, as a function of 
look-back time and for delay times of 2, 4.5, and 6~Gyr (dashed, solid, and dotted lines, respectively). 
The solid line corresponds to our own case. 
}
\label{fig:civ}
\end{figure}

On the other hand, the probabilities of the emergence of life, of Earth-like biospheres, and of 
intelligent species are so far unconstrained. Consequently, we can only estimate the frequency of 
civilizations with respect to some arbitrary reference point. Here we choose to normalize the number of 
planets estimated with the modified criterion to 1 at $z=0$~for MW-type disks. 
We then find that, if the number of Earth-like biospheres or civilizations now present in the MW 
is 1, a typical $10^{11}$~M$_{\odot}$~galaxy has $\sim$2-3, and $10^{9-0.22\times(t_{\text{min}}/Gyr)}$~civilizations 
exist in the observable universe when accounting for look-back time. 
The occurrence rate of civilizations, as shown in Fig.~\ref{fig:civ}, peaks $3+1.3\times t_{\text{min}}$~Gyr after 
the onset of star formation. This suggests that, if Earth is not unique in this galaxy and the timescale 
of our own evolution is typical, we exist $\sim$4~Gyr after this maximum, with the median age of civilizations 
being $\sim$3.5~Gyr. Conversely, assuming that $t_{\text{min}}=4.5$~Gyr, $1.4\times10^9$~planets following the 
above criterion have formed in a MW-mass disk since the onset of star formation. 
For us to exist before most civilizations our galaxy will produce would then imply either that the incidence 
of civilizations per suitable planet is $<7\times10^{-10}$~or that, contrary to what the timescale of our 
evolution suggests, the typical delay time is $>8$~Gyr.

\section{\label{conc}Conclusion}

We have used an analytic model of galaxy evolution to estimate galaxy habitability, measured by the fraction 
of stars with habitable planets, as a function of galaxy type, M$_{\star}$, and redshift. Our model includes 
passive galaxies through a simple treatment of galaxy quenching and metal enrichment, as well as thermal 
effects of stellar evolution and supernovae on habitable zone planets. We summarize our findings as follows:\\

$\bullet$~We consider two different types of metallicity dependence for the frequency of terrestrial planets 
and find that, given the assumptions of the model, a weak negative metallicity dependence (Case 2) reproduces 
observations better. In this case, between 0.65\% and 0.8\% of stars in $>10^9$~M$_{\odot}$~galaxies 
are expected to host planets in their HZ, close to the $<1$\% inferred from \emph{Kepler} observations. 
The habitability of galaxies at $z=0$~varies then relatively little with galaxy mass, reaching a maximum at 
$\log M_{\star}=10.6$.\\

$\bullet$~We estimate that the radius where the thermal effects of SN on planets become significant is 
0.3-0.5~pc. 
The impact of SN on galaxy habitability is therefore negligible, as the fractional irradiated volume is 
almost always very small except at high M$_{\star}$~and redshift, where SFRs are extreme and 
galaxies more compact. The effect of the central AGN is likewise limited.\\

$\bullet$~The habitability of passive galaxies is slightly but systematically higher than that of 
star-forming galaxies and has remained mostly unchanged since $z\sim1.5$. On the other hand, the habitability 
of SF galaxies has increased monotonically since $z=2$~and has only recently started to 
plateau. Overall, the average habitability of galaxies has not varied much in the last 8~Gyr.\\

$\bullet$~The median age of habitable planets is $\sim$6~Gyr in SF galaxies, i.e., $\sim$1.5~Gyr 
older than our own solar system, and $\sim$10~Gyr in passive galaxies. Using a more restrictive criterion, 
the occurrence rate of habitable planets similar to present-day Earth (which we can assume is proportional 
to that of alien civilizations) peaked $\sim$4~Gyr ago. For us to be unique in the history of the MW would 
imply that the probability for intelligent life to evolve on a suitable planet is lower than 
$\sim$7$\times10^{-10}$.\\

\begin{acknowledgements}

We thank E. Daddi for enlightening discussions, F. Adams, C. Park, O. Snaith, and H.S. Hwang for their 
helpful suggestions which helped improve this paper.
This research has made use of the NASA Exoplanet Archive, which is operated by the California Institute 
of Technology, under contract with the National Aeronautics and Space Administration under the Exoplanet 
Exploration Program.

\end{acknowledgements}

\end{document}